\documentclass[aps,preprint]{revtex4}%
\usepackage{amsfonts}
\usepackage{amsmath}
\usepackage{amssymb}
\usepackage{graphicx}%
\providecommand{\U}[1]{\protect\rule{.1in}{.1in}}

\begin{document}
\title{Dynamics of the one-dimensional random transverse Ising model with
next-nearest-neighbor interactions}
\author{Xiao-Juan Yuan}
\author{Xiang-Mu Kong}
\thanks{Corresponding author.}
\email{kongxm@mail.qfnu.edu.cn (X.-M. Kong)}
\affiliation{College of Physics and Engineering, Qufu Normal University, Qufu 273165, China}
\author{Zhong-Qiang Liu}
\affiliation{Qindao College, Qingdao Technological University, Qingdao 266106, China}
\author{Zhen-Bo Xu}
\affiliation{College of Physics and Engineering, Qufu Normal University, Qufu 273165, China}

\begin{abstract}
The dynamics of the one-dimensional random transverse Ising model with both
nearest-neighbor (NN) and next-nearest-neighbor (NNN) interactions is studied
in the high-temperature limit by the method of recurrence relations. Both the
time-dependent transverse correlation function and the corresponding spectral
density are calculated for two typical disordered states. We find that for the
bimodal disorder the dynamics of the system undergoes a crossover from a
collective-mode behavior to a central-peak one and for the Gaussian disorder
the dynamics is complex. For both cases, it is found that the central-peak
behavior becomes more obvious and the collective-mode behavior becomes weaker
as $K_{i}$ increase, especially when $K_{i}>J_{i}/2$ ($J_{i}$ and $K_{i}$ are
exchange couplings of the NN and NNN interactions, respectively). However, the
effects are small when the NNN interactions are weak ($K_{i}<J_{i}/2$).

\end{abstract}
\maketitle

\section{Introduction\label{introduction}}

There has been a considerable interest in the study of the dynamics of quantum
spin systems in the past few decades\cite{1,2,book}, and the calculation of
dynamic correlation functions remains a highly nontrivial and real challenging
task until now. Some exact results have been obtained for the one-dimensional
(1-D) pure quantum spin models, e.g., the 1-D transverse Ising model and the
1-D $XY$ model\cite{3,4,5}. Recently, more attention has been paid to the
investigation of the dynamical behavior of disordered
systems\cite{disorderbook,Florencio1999,Liu2006,xy,xy2,xy1}, which can be
applied to describe the properties of many materials such as window glass,
magnets with frozen-in disorder, etc.. One simple but important example of
such systems is the 1-D random transverse Ising model (RTIM).

The dynamical behavior of the 1-D RTIM with the bimodal distribution is
studied by Florencio and Barroto\cite{Florencio1999}, and it is found that the
dynamics undergoes a crossover from a central peak behavior onto a collective
mode one. Recently, we have investigated the effects of Gaussian disorder on
the dynamics of the 1-D RTIM\cite{Liu2006}, and have found that there are two
crossovers when the standard deviation of random variables is small and there
is no crossover if the value of the standard deviation is large enough.
Besides, the dynamical behavior of the random-bond transverse Ising model with
four-spin interactions\cite{fourspin} and the disordered $XY$
chain\cite{xy1,xy,xy2} have been studied.

In the work mentioned above only nearest-neighbor (NN) interactions are
considered. To our knowledge, no related results of disordered quantum spin
systems with next-nearest-neighbor (NNN) interactions have been reported so
far. However, the work of Sen has shown the role of second neighbor
interactions on the relaxation in $s=1/2$ pure quantum spin
chains\cite{Sen1996}. The results show that the dynamical correlation in the
NNN transverse Ising chain is noticeably different with that of the exactly
solvable NN transverse Ising chain. Therefore, it is expected that considering
additional spin-spin interactions in disordered systems will make some
differences in the dynamical process.

Our main interest is to investigate the effects of NNN interactions on the
dynamics of the 1-D disordered quantum spin systems. It is well known that the
interactions between spins may be complex in real materials. For studying the
properties of real systems theoretically and experimentally, the easy way is
to consider a model with only the dominant NN interactions. In this paper, we
consider the 1-D RTIM with both NN and NNN interactions. The 1-D RTIM can be
used to describe the dynamical property of many condensed-matter systems like
the quasi-one-dimensional ferroelectric crystals (e.g., PbH$_{1-x}$D$_{x}%
$Po$_{4}$)\cite{Plascak1982,Levitsky1986}, and the Ising spin glass
LiHo$_{0.167}$Y$_{0.833}$F$_{4}$\cite{Wu1991}. We investigate the cases that
the exchange couplings between spins or transverse fields independently
satisfy the bimodal distribution and the Gaussian distribution, respectively.
Our calculations are based on the method of recurrence
relations\cite{recursion,recursion1} which is very powerful in the study of
classical and quantum many-body dynamics\cite{book,5,harmonicchain}.
Meanwhile, we also used some reliable approximation schemes such as the
so-called Gaussian terminator\cite{gaussian1,gaussian,book} and the Pad\'{e}
approximants. It is found in both disorder that the central-peak behavior
becomes more obvious and the collective-mode behavior becomes weaker when
$K_{i}>J_{i}/2$. We also find that the dynamics of the system is not sensitive
to the property of the NNN interaction whether it is ferromagnetic or antiferromagnetic.

This paper is arranged as follows. In Sec. \ref{model method} we introduce the
model used in this paper and the method of recurrence relations. Secs.
\ref{bimodal disorder} and \ref{gaussian disorder} give the dynamical results
for the bimodal disorder and the Gaussian disorder, respectively. Sec.
\ref{conclusions} provides conclusions.

\section{Model and method\label{model method}}

The Hamiltonian of the 1-D RTIM with both NN and NNN interactions can be
written as%
\begin{equation}
H=-\frac{1}{2}%
{\displaystyle\sum\limits_{i}^{N}}
\left(  J_{i}\sigma_{i}^{x}\sigma_{i+1}^{x}+K_{i}\sigma_{i}^{x}\sigma
_{i+2}^{x}\right)  -\frac{1}{2}%
{\displaystyle\sum\limits_{i}^{N}}
B_{i}\sigma_{i}^{z}, \label{H}%
\end{equation}
where $\sigma_{i}^{\alpha}$ ($\alpha=x,y,z$) are Pauli matrices at site $i$,
$B_{i}$ denote the external fields, while $J_{i}$ and $K_{i}$ are exchange
couplings between NN spins and NNN spins, respectively. The periodic boundary
conditions $\sigma_{i+N}^{\alpha}=\sigma_{i}^{\alpha}$ are assumed in next
calculation, where $N$ is the number of spins. For simplicity, we assume that
$K_{i}=\alpha J_{i}$ ($0\leqslant\alpha<1$) and consider $J_{i}$ and $B_{i}$
are uncorrelated random variables which satisfy the probability distributions
$\rho(J_{i})$ and $\rho(B_{i})$, respectively. It is obvious that, in the
limit $K_{i}\rightarrow0$ in Eq. (\ref{H}), this model can be reduced to the
1-D RTIM\cite{Florencio1999,Liu2006}.

The dynamical behavior of classical or quantum many-body systems is
conveniently expressed in terms of dynamic correlation functions. In this
paper, we are interested in the time-dependent transverse correlation function
defined by%
\begin{equation}
C\left(  t\right)  =\overline{\left\langle \sigma_{j}^{x}\left(  t\right)
\sigma_{j}^{x}\left(  0\right)  \right\rangle }, \label{Ct}%
\end{equation}
where $\overline{\left\langle \cdots\right\rangle }$ denotes an ensemble
average followed by an average over the disorder variables. The spectral
density $\Phi\left(  \omega\right)  $ ($\omega$ is the frequency) which is
able to be determined directly from experiments is defined as the Fourier
transformation of the correlation function,%
\begin{equation}
\Phi\left(  \omega\right)  =\int_{-\infty}^{+\infty}e^{i\omega t}C\left(
t\right)  dt. \label{spectraldensity}%
\end{equation}

The method of recurrence relations has already been applied to solve a variety
of many-body systems, such as the classical harmonic chain\cite{harmonicchain}%
, the electron gas\cite{recursion,recursion1}, spin
systems\cite{gaussian,sen1991,v.s.} and ergodic theory\cite{ergodic1} etc.,
successfully. In the following we will summarize this method.

Consider a many-body system defined by a Hamiltonian $H$. The time evolution
of a dynamical operator $A$ is described by the Liouville (or Heisenberg)
equation of motion%
\begin{equation}
\frac{dA\left(  t\right)  }{dt}=iLA\left(  t\right)  , \label{liouville}%
\end{equation}
where $L$ is the Liouville operator, $LA=\left[  H,A\right]  \equiv HA-AH$.
The solution of Eq. (\ref{liouville}) can be given as the form of the
orthogonal expansion\cite{recursion}%
\begin{equation}
A(t)=\sum_{\nu=0}^{\infty}a_{\nu}\left(  t\right)  f_{\nu}, \label{At}%
\end{equation}
where $\left\{  f_{\nu}\right\}  $ are an orthogonal set of basis vectors
spanning a Hilbert space $S$, the coefficients $a_{\nu}\left(  t\right)  $ are
time dependent functions representing the projection of $A(t)$ onto $f_{\nu}$
at $t$.

In the high-temperature limit $T=\infty$, the inner product which includes
both the statistical and random averages in our system is described
as\cite{Florencio1999,xy,xy1}%
\begin{equation}
\left(  X,Y\right)  =\overline{\langle XY^{\dagger}\rangle},
\label{innerproduct}%
\end{equation}
where $X$ and $Y$ are basis vectors defined in $S$.

Set $f_{0}=A\left(  0\right)  $, which gives $a_{0}\left(  0\right)  =1$ and
$a_{\nu}\left(  0\right)  =0$ for $\nu>0$ by Eq. (\ref{At}). The basis vectors
$f_{\nu}$ satisfy the recurrence relation (RRI)%
\begin{equation}
f_{\nu+1}=iLf_{\nu}+\Delta_{\nu}f_{\nu-1},\quad\nu\geqslant0, \label{RRI}%
\end{equation}
where the coefficients, also known as recurrants, are defined as%
\begin{equation}
\Delta_{\nu}=\frac{\left(  f_{\nu},f_{\nu}\right)  }{\left(  f_{\nu-1}%
,f_{\nu-1}\right)  }\quad(\nu\geqslant1) \label{recurrant}%
\end{equation}
with $f_{-1}\equiv0$ and $\Delta_{0}\equiv1.$ Meanwhile, the coefficients
$a_{\nu}\left(  t\right)  $ satisfy a second recurrence relation (RRII)%
\begin{equation}
\Delta_{\nu+1}a_{\nu+1}\left(  t\right)  =-\frac{da_{\nu}\left(  t\right)
}{dt}+a_{\nu-1}\left(  t\right)  ,\quad\nu\geqslant0, \label{RRII}%
\end{equation}
where $a_{-1}\left(  t\right)  \equiv0$, and $a_{0}\left(  t\right)  $ is the
time-dependent correlation function $\left\langle A(t)A(0)\right\rangle $.
Obviously, by choosing $f_{0}=\sigma_{j}^{x}$ the average spin correlation
function is just given by $C\left(  t\right)  =\overline{\left\langle
\sigma_{j}^{x}\left(  t\right)  \sigma_{j}^{x}\left(  0\right)  \right\rangle
}$ (Eq. (\ref{Ct})), which can be written as the form of moment expansion
\[
C\left(  t\right)  =\sum_{k=0}^{\infty}\frac{\left(  -1\right)  ^{k}}{\left(
2k\right)  !}\mu_{2k}t^{2k}%
\]
with%
\begin{equation}
\mu_{2k}=\frac{1}{Z}\overline{\text{Tr}\sigma_{j}^{x}\left[  H,\left[
H,\cdots\left[  H,\sigma_{j}^{x}\right]  \cdots\right]  \right]  },
\label{ctm2}%
\end{equation}
where $\mu_{2k}$ is the $2k$th moment of $C\left(  t\right)  $. The partition
function $Z=$Tr$1=2^{N}$ equals the number of quantum states of the system.
Using the first $2\nu$ moments, we can calculate the correlation function
$C\left(  t\right)  $ by constructing Pad\'{e} approximants.

By taking the Laplace transformation of the recurrence relation (RRII), one
obtains%
\begin{equation}
\Delta_{\nu+1}a_{\nu+1}\left(  z\right)  -\delta_{\nu,0}=-za_{\nu}\left(
z\right)  +a_{\nu-1}\left(  z\right)  ,\quad\nu=0,1,2\cdots, \label{az1}%
\end{equation}
where $z=\varepsilon+i\omega$ ($\varepsilon>0$) is a variable of the complex
plane. Then one can get the continued-fraction form%
\begin{equation}
a_{0}\left(  z\right)
=\cfrac{1}{z+\cfrac{\Delta_{1}}{z+\cfrac{\Delta_{2}}{z+\cdots}}}. \label{az2}%
\end{equation}
Furthermore, it is proved that the spectral density $\Phi\left(
\omega\right)  $ (Eq. (\ref{spectraldensity})) is able to be determined
directly by Eq. (\ref{az2}),%
\begin{equation}
\Phi\left(  \omega\right)  =\lim_{\varepsilon\rightarrow0}\operatorname{Re}%
a_{0}\left(  z\right)  . \label{specdensity1}%
\end{equation}
Note that $\Delta_{\nu}$ are the key quantities for calculating the dynamic
correlation functions.

Generally, only a finite number of continued-fraction coefficients can be
determined. So it is necessary to use a scheme to terminate the continued
fraction. The one that serves our model best is the so-called Gaussian
terminator\cite{book,gaussian1,gaussian}. Suppose the first $M$ recurrants are
determined, in this approximation, the others are assumed to be of the form
$\Delta_{\nu}=\nu\left(  \Delta_{M}/M\right)  $, for $\nu>M$.

\section{Dynamics for bimodal disorder\label{bimodal disorder}}

After a lengthy calculation, the first eight basis vectors are exactly
obtained by Eq. (\ref{RRI}). In the following, we just give the first two of
them:%
\begin{align}
f_{1}  &  =B_{j}\sigma_{j}^{y},\nonumber\\
f_{2}  &  =\left(  \Delta_{1}-B_{j}^{2}\right)  \sigma_{j}^{x}+B_{j}%
K_{j-2}\sigma_{j-2}^{x}\sigma_{j}^{z}+B_{j}J_{j-1}\sigma_{j-1}^{x}\sigma
_{j}^{z}+\nonumber\\
&  B_{j}J_{j}\sigma_{j+1}^{x}\sigma_{j}^{z}+B_{j}K_{j}\sigma_{j+2}^{x}%
\sigma_{j}^{z}.
\end{align}
The squared norms of the basis vectors are given by Eq. (\ref{innerproduct})
as follows:%
\begin{align}
\left(  f_{0},f_{0}\right)   &  =1,\nonumber\\
\left(  f_{1},f_{1}\right)   &  =\overline{B_{j}^{2}},\nonumber\\
\left(  f_{2},f_{2}\right)   &  =\overline{\Delta_{1}^{2}}-2\overline
{\Delta_{1}}\overline{B_{j}^{2}}+\overline{B_{j}^{4}}+\overline{B_{j}^{2}%
}\overline{J_{j-1}^{2}}+\overline{B_{j}^{2}}\overline{J_{j}^{2}}%
+\overline{B_{j}^{2}}\overline{K_{j-2}^{2}}+\overline{B_{j}^{2}}%
\overline{K_{j}^{2}}.
\end{align}
Using the above results, we have calculated the first eight coefficients
$\Delta_{1},$ $\Delta_{2},$ $\cdots,$ and $\Delta_{8}$ exactly, and the
$\Delta_{9}$ through the assumption $\Delta_{\nu}=\nu\left(  \Delta
_{M}/M\right)  $ approximately. Meanwhile, the first 18 moments are obtained
and the correlation function $C\left(  t\right)  $ can be determined by
constructing the Pad\'{e} approximants.

Notice that the coefficients (see Eq. (\ref{recurrant})) are even functions of
$K_{i}$ and the correlation functions are determined uniquely by the
recurrants. Thus, the dynamical property of the system is independent of that
the NNN interactions are ferromagnetic or antiferromagnetic. Actually, the
system is in its paramagnetic phase in the high-temperature limit. Next, we
only consider the case of ferromagnetic NNN interactions ($K_{i}>0$). We
calculate two typical cases that the random variables satisfy the bimodal
distribution and the Gaussian distribution, respectively.

In the following, we assume that the exchange couplings $J_{i}$ or the
transverse fields $B_{i}$ satisfy the bimodal distribution
\begin{equation}
\rho\left(  \left\{  \beta_{i}\right\}  \right)  =%
{\textstyle\prod\limits_{i}^{N}}
[p\delta\left(  \beta_{i}-\beta_{a}\right)  +\left(  1-p\right)  \delta\left(
\beta_{i}-\beta_{b}\right)  ], \label{b1}%
\end{equation}
where $\beta_{i}=J_{i}$ or $B_{i}$, $p$ is the concentration of coupling
$J_{a}$ or magnetic field $B_{a}$ and takes values from 0 to 1.

We first consider the random band and uniform field model. In this case,
without loss of generality we set $B_{i}=B=1$, and choose $J_{a}=1.0$ and
$J_{b}=0.4$, which have been used in Ref. \cite{Florencio1999}. In this
assumption, the exchange couplings $J_{i}$ change from $J_{i}<$ $B$ ($p<1$) to
$J_{i}=$ $B$ ($p=1$). The transverse correlation functions $C\left(  t\right)
$ and the corresponding spectral densities $\Phi\left(  \omega\right)  $ are
given in Fig. 1 for several values of bond concentration $p$. In order to show
better the effects of the NNN interactions on the dynamics of the system, we
have considered the cases that $K_{i}=0$, $J_{i}/4$, $J_{i}/2$ and $3J_{i}/4$,
respectively. Obviously, for the $K_{i}=0$ case, the results are just of the
1-D RTIM studied by Florencio and Barreto\cite{Florencio1999}.%

\begin{figure}
[ptb]
\begin{center}
\includegraphics[scale=1.3]{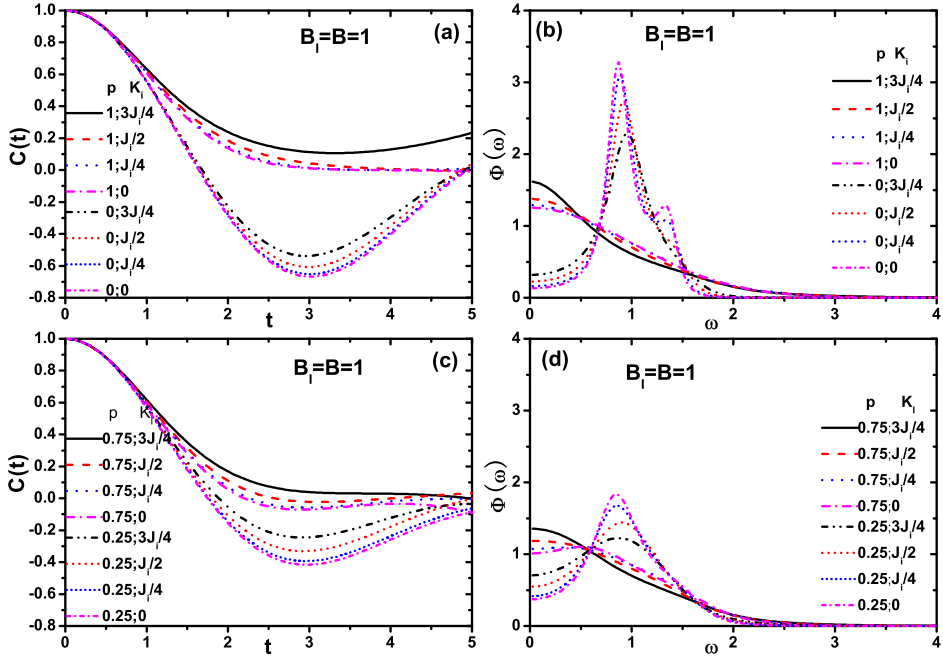}
\caption{Time-dependent correlation functions $C\left( t\right)  $
and corresponding spectral densities $\Phi\left( \omega\right)  $
for the case of random bond which satisfy the bimodal distribution
in which $J_{a}=$ $1.0$ and$\ J_{b}=0.4$. (a) and (b) plot of the
pure cases that $p=0$ and $1$. The results in (c) and (d) are the
disordered cases for $p=0.25$ and $0.75$. The central-peak behavior
becomes more obvious and the collective-mode behavior becomes weaker
as $K_{i}$ increase. The black solid line in (a) is monotonic
for $t<3$.}%
\end{center}
\end{figure}

From different cases of $K_{i}$ shown in Fig. 1, we can see that the system
shows a collective-mode behavior for small values of $p$ (i.e. $p=0$ or
$0.25$), and exhibits a central-peak behavior when $p=1$. In general, when the
external field is small, the spin-spin interactions play an important role,
thus the central-peak behavior dominates the dynamics of the system, while for
large $B$ the dynamical behavior is collective-mode one, which is due to the
precession of spins in the transverse field. This means that the dynamics of
the present model with bimodal distributions is similar to that of Ref.
\cite{Florencio1999}.

By comparing the curves for the cases that $K_{i}=J_{i}/4$, $J_{i}/2$, and
$3J_{i}/4$ with those of the $K_{i}=0$ case (see Fig. 1), we can see that the
dynamics has no evident change if the NNN interactions are weak (e.g.
$K_{i}=J_{i}/4$). However, there are some obvious differences when
$K_{i}>J_{i}/2$. The dot-dashed curve for the pure case $p=1$ when $K_{i}=0$
in Fig. 1(a) describes the dynamics for the exactly solvable limit in which
$J_{i}=B_{i}=1$, now the $C\left(  t\right)  $ is a Gaussian function\cite{5}.
Meanwhile, the other curves for the cases that $p=1$ when $K_{i}=J_{i}/4$,
$J_{i}/2$, and $3J_{i}/4$ in Fig. 1(a), respectively, all behave monotonically
but exhibit slower decay than for the $K_{i}=0$ case, and are not a Gaussian.
On the other hand, the lines shown in Fig. 1(a) for the pure case $p=0$
indicate that the collective-mode behavior becomes weaker as $K_{i}$ increase.
The same results can be also obtained from the corresponding spectral
densities. As shown in Fig. 1(b) the central peak increases, meanwhile, the
collective-mode peak becomes lower and the width of the spectral line broaden
as $K_{i}$ increase. For the disordered case that $p=0.25$ (see Figs. 1(c) and
(d)), we also find weaker collective-mode behavior if the NNN interactions
become stronger. However, for the disordered case that $p=0.75$ when
$K_{i}=3J_{i}/4$ (see the black solid line in Fig. 1(c) ), the $C\left(
t\right)  $ decays monotonically to zero, and the dynamics of the system is a
central-peak behavior which is not as the case that $p=0.75$ when $K_{i}=0$
(the lines for $p=0.75$ when $K_{i}=0$ or $J_{i}/4$ in Figs. 1(c) and (d) show
a disordered behavior which is something between the collective-mode behavior
and the central-peak one). All the above results indicate that the
interactions between spins are stronger in our system, and that the effects of
the NNN interactions on the dynamics of the system cannot be neglected.

We now consider the random field and uniform band model, in which the
transverse fields satisfy the bimodal distribution $\rho\left(  B_{i}\right)
$ and can take the values $B_{a}=0.6$ ($p=1$) and $B_{b}=1.4$ ($p=0$), while
the exchange couplings are constants ($J_{i}=J=1$, $K_{i}=0$, $1/4$, $1/2$ or
$3/4$). This allows the external fields changing from $B_{i}>J$ to $B_{i}<J$
as $p$ increases. The results of $C\left(  t\right)  $ and $\Phi\left(
\omega\right)  $ for different values of $p$ are shown in Fig. 2. The curves
for $p=0$ are the pure cases dominated by the stronger field energy. In this
case, the system is at the collective mode regime. When $p=1$ and $0.75$, the
correlation functions decay monotonically, and thus the dynamics is dominated
by the central-peak behavior. However, for the disordered case $p=0.25$, the
dynamics of the system is neither central-peak nor collective-mode type, but
something between them. Hence, for this model, the system also undergoes a
crossover from a collective-mode behavior to a central-peak one as $p$
increases from $0$ to $1$.%

\begin{figure}
[ptb]
\begin{center}
\includegraphics[scale=1.3]{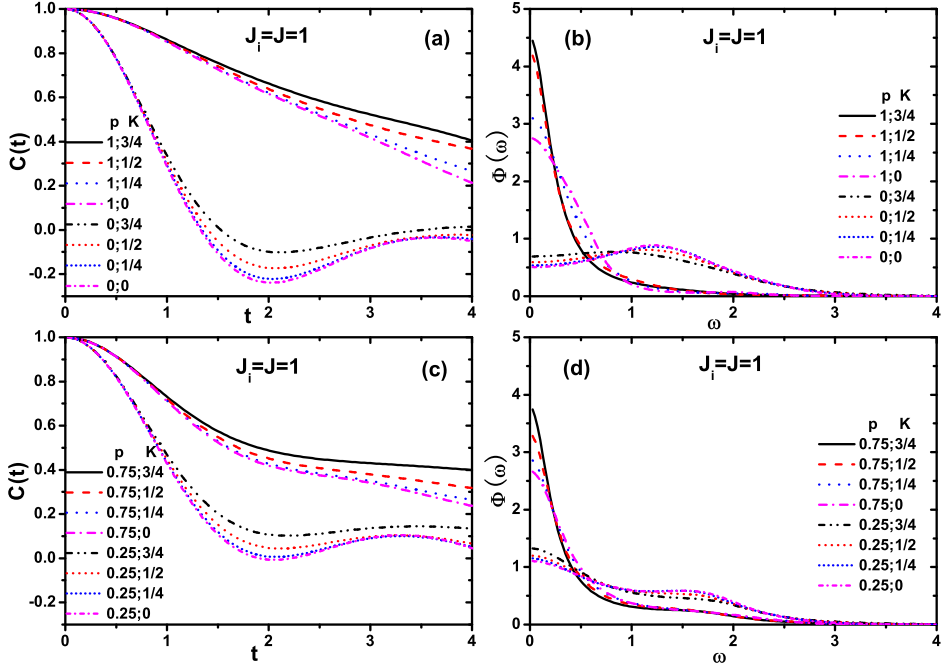}
\caption{Correlation functions and corresponding spectral densities
for the random field model, in which $B_{a}=0.6$ and $B_{b}=1.4$.
The system undergoes a crossover from a collective-mode behavior to
a central-peak one as $q$ increases. The central-peak behavior
becomes more obvious and the
collective-mode behavior becomes weaker as $K_{i}$ change from 0 to 3/4.}%
\end{center}
\end{figure}

The same as the above random bond model, it is found that the central-peak
behavior becomes more obvious and the collective-mode behavior becomes weaker
as $K_{i}$ increase, especially when $K_{i}>J_{i}/2$. From the curves for
$p=1$ in Fig. 2(a) and $p=0.75$ in Fig. 2(c), we can find that the $C\left(
t\right)  $ decays more slowly as $K_{i}$ changing from $0$ to $3/4$.
Meanwhile, the magnitude for $\Phi\left(  \omega\right)  $ at $\omega=0$
increases as $K_{i}\neq0$ (see Figs. 2(b) and (d)). The results of $p=0$ in
Fig. 2(a) and (b) indicate that the collective-mode behavior becomes weaker if
the NNN interactions are stronger, since the oscillatory curves are less damped.

\section{Dynamics for Gaussian disorder\label{gaussian disorder}}

In the following, we assume that the exchange couplings $J_{i}$ or the
transverse fields $B_{i}$ are uncorrelated random variables which satisfy the
Gaussian distribution\cite{Liu2006,gaussiandistribution}%
\begin{equation}
\rho\left(  \left\{  \beta_{i}\right\}  \right)  =\prod\limits_{i}^{N}\frac
{1}{\sqrt{2\pi}\sigma_{\beta}}\exp\left[  -\frac{\left(  \beta_{i}%
-\beta\right)  ^{2}}{2\sigma_{\beta}^{2}}\right]  ,\label{gd1}%
\end{equation}
where $\beta$ denotes the mean value of the random variables $\beta_{i}$, and
$\sigma_{\beta}$ is the standard deviation. Next, we discuss two different
cases that the random-bond and the random-field model, respectively. We find
that the effects of $K_{i}$ on the dynamics are not obvious when $K_{i}<$
$J_{i}/2$, which is similar as the above case of the bimodal disorder. In the
following, we only give the results for $K_{i}=J_{i}/2$ and $3J_{i}/4$ in the
random-bond model, and for $K_{i}=3J_{i}/4$ in the random-field model.%

\begin{figure}
[ptb]
\begin{center}
\includegraphics[scale=1.3]{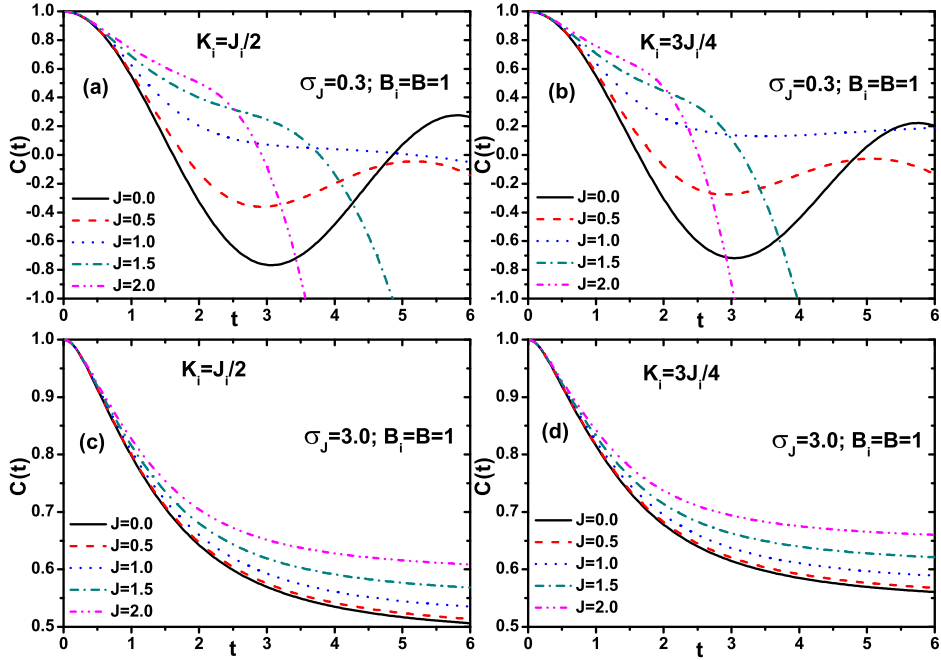}
\caption{Correlation functions $C\left(  t\right)  $ for the case
that the bonds satisfy the Gaussian distribution while the external
fields are constants. Two typical cases that $\sigma_{J}=0.3$ and
$3.0$ are considered.
The mean value $J$ varies from $0$ to $2$.}%
\end{center}
\end{figure}
%

\begin{figure}
[ptb]
\begin{center}
\includegraphics[scale=1.3]{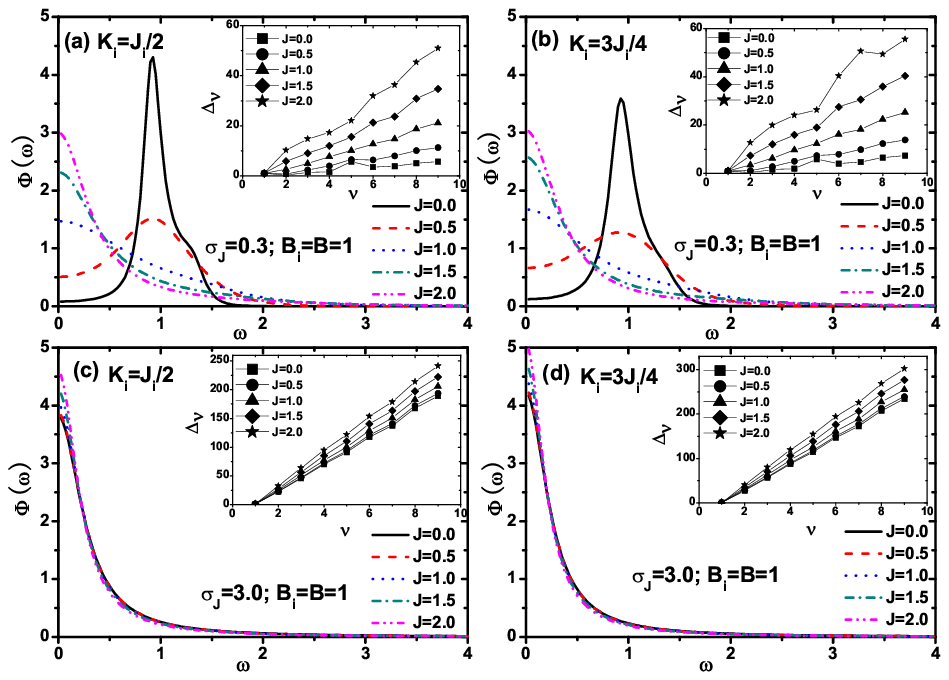}
\caption{Corresponding spectral densities $\Phi\left(  \omega\right)
$ for the same parameters as in Fig. 3. For $\sigma_{J}=3.0$ the
central-peak increases as $K_{i}$ change from $J_{i}/2$ to
$3J_{i}/4$. For $\sigma_{J}=0.3$
the collective-mode behavior becomes weaker as $K_{i}$ increase.}%
\end{center}
\end{figure}

For the random-bond model, the exchange couplings $J_{i}$ satisfy the Gaussian
distribution while the transverse fields $B_{i}$ are constants. We keep
$B_{i}=B=1$ which sets the energy scale, and consider that the mean value $J$
varies from $0$ to $2$ and the standard deviation $\sigma_{J}$ changes from
$0.3$ to $3.0$. Both the correlation functions $C\left(  t\right)  $ and the
corresponding spectral densities $\Phi\left(  \omega\right)  $ are shown in
Figs. 3 and 4, respectively. The insets to Fig. 4 present the first nine recurrants.

From Figs. 3 and 4 we can see that for the case of small values of $\sigma
_{J}$ (e.g., 0.3), there are two typical dynamics: the collective-mode
behavior and the central-peak behavior. It is obvious that the black solid
curve for $J=0$ in Fig. 3(a) is a damped cosine function, which is due to the
precession of spins in an external transverse magnetic field\cite{Liu2006}. As
$J$ increases, the system first shows a weak collective-mode behavior for the
case of $J<B$ (e.g., $J=0.5$), then exhibits a central-peak behavior when
$J>B$ (e.g., $J=1.5$).

For the case of large $\sigma_{J}$ (e.g., 3.0), it is found that the system
only shows a central-peak behavior, and there is no crossover. In this case,
the strong exchange couplings play an important role in the dynamics of the
system. That is, the spin-spin interactions are dominant in the competition
between the spin-spin interactions and the external fields. It is also found
that, further increasing the NNN interactions will make the curves of
$C\left(  t\right)  $ for the central-peak behavior decay more slower and the
magnitude for $\Phi\left(  \omega\right)  $ at $\omega=0$ become larger. This
all indicate that the dynamical behavior of the system is sensitive to the
inclusion of the NNN interactions.%

\begin{figure}
[ptb]
\begin{center}
\includegraphics[scale=1.3]{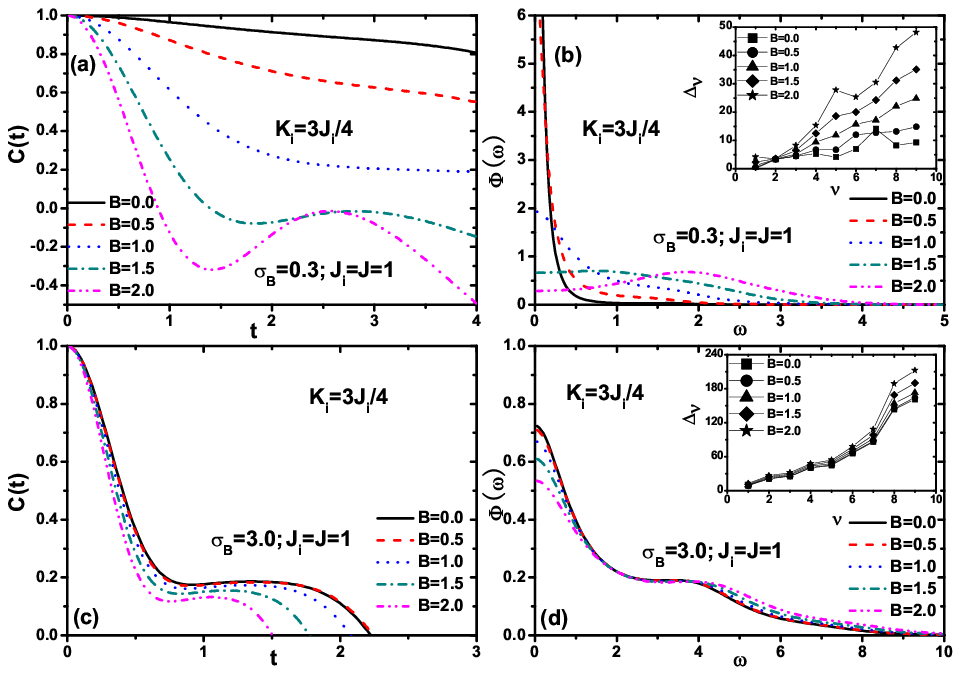}
\caption{Correlation functions and corresponding spectral densities
for the random field model when $K_{i}=3J_{i}/4$. The insets to (b)
and (d) present the first nine recurrants. Let the mean value $B$
varies from $0$ to $2$. For $\sigma_{B}=0.3$ there is a crossover
from a central-peak behavior to a collective-mode one as $B$
increases. For $\sigma_{B}=3.0$ the system exhibits
a most disordered behavior.}%
\end{center}
\end{figure}

We next discuss the results of the random-field model, in which the transverse
fields $B_{i}$ satisfy the Gaussian distribution, while the exchange couplings
remain unaltered ($J_{i}=J=1$, $K_{i}=3/4$). Let the mean value $B$ varies
from 0 to 2. From Fig. 5 we can see that for $\sigma_{B}=0.3$ the system
undergoes a crossover from a central-peak behavior to a collective-mode one as
$B$ increases. When $\sigma_{B}$ is large enough ($\sigma_{B}=3.0$), the
system only shows one type of dynamics and there is no crossover. That is a
most-disordered state\cite{Liu2006}, which is something between the
central-peak behavior and the collective-mode behavior. By comparing the
results with those of Ref. \cite{Liu2006}, we can find that the oscillatory
behavior becomes weaker as $K_{i}$ increase. However, different from the
effects of the NNN interactions on the dynamics of the random-bond model,
increasing $K_{i}$ in this case will make no difference when $\sigma_{B}$ is
large enough. That is, the dynamics of the system now is dominated by the
disordered external field.

\section{Conclusions\label{conclusions}}

In this paper, we have studied the effects of the NNN interactions on the
dynamics for the 1-D RTIM in the high-temperature limit. We have considered
the cases that the random variables satisfy the bimodal distribution and the
Gaussian distribution, respectively. It is found in both cases that the
dynamical property of the present model is similar to that of the 1-D RTIM
when $K_{i}<J_{i}/2$. However, the central-peak behavior becomes more obvious
and the collective-mode behavior becomes weaker as the NNN interactions
increase (i.e. $K_{i}>J_{i}/2$). It is expected that we can get similar
results in other disordered quantum spin systems, e.g., the $XY$ and the $XYZ$ models.

\begin{acknowledgments}
This work is supported by the National Natural Science Foundation of China
under Grant No. 10775088, the Shandong Natural Science Foundation under Grant
No. Y2006A05, and the Science Foundation of Qufu Normal University. One of the
authors (Yuan) thanks Jia-Xue Liu, Ying Wang and Wei-Ke Zou for valuable discussions.
\end{acknowledgments}

\newpage

\end{document}